\documentclass[12pt]{article}
\usepackage{amsmath,amssymb}
\usepackage{graphicx,color}
\numberwithin{equation}{section}
\usepackage{cite}
\usepackage{bm}
\usepackage{dcolumn}  
\newcommand{\beq}{\begin{equation}}
\newcommand{\eeq}{\end{equation}}
\newcommand{\beqa}{\begin{eqnarray}}
\newcommand{\eeqa}{\end{eqnarray}}
\newcommand{\beqar}{\begin{eqnarray*}}
\newcommand{\eeqar}{\end{eqnarray*}}
\newcommand{\al}{\alpha}

\newcommand{\eg}{{\it e.g.,}\ }
\newcommand{\ie}{{\it i.e.,}\ }
\newcommand{\labell}[1]{\label{#1}} %{\label{#1}} %
\newcommand{\reef}[1]{(\ref{#1})}
\newcommand\prt{\partial}

\newcommand\cF{{\cal F}}

\begin{document}
\baselineskip 18pt%
\begin{titlepage}
\vspace*{1mm}%
\hfill%
\vspace*{15mm}%

\centerline{{\Large {\bf More on  entropy function formalism  }}}
\centerline{{\Large{\bf   for  non-extremal branes}}}
\vspace*{5mm}
\begin{center}
{ Mohammad R. Garousi, Ahmad Ghodsi, Tooraj Houri,  Mehran Khosravi }\\
\vspace*{0.2cm}
{ Department of Physics, Ferdowsi University of Mashhad, \\
P.O. Box 1436, Mashhad, Iran}\\
\vspace*{0.1cm}
%and\\
{ School of Physics, 
 Institute for research in fundamental sciences (IPM), \\
 P.O. Box 19395-5531, Tehran, Iran. 
}\\\vspace*{1.5cm}
\end{center}

\begin{abstract} 
We find $R^4$ correction to the non-extremal $D1D5P$  solution of the supergravity by exactly solving the differential equations of motion and by using the entropy function formalism. In both cases, we find the same correction to the supergravity solution. We then calculate the correction to the entropy using the free energy method and the entropy function formalism. The results are the same.
\end{abstract} 

\end{titlepage}

%%%%%%%%%%%%%%%%%%%%%%%%%%%%%%%%%%%%%%%%%%%%%%%%%%%%%%%%%%%%%%%%%%%%%%%%%
%%%%%%%%%%%%%%%%%%%%%%%%%%%%%%%%%%%%%%%%%%%%%%%%%%%%%%%%%%%%%%%%%%%%%%%%%
%%%%%%%%%%%%%%%%%%%%%%%%%%%%%%%%%%%%%%%%%%%%%%%%%%%%%%%%%%%%%%%%%%%%%%%%% 

\section{Introduction} \label{intro} 

One way of calculating the entropy of a black hole in higher derivative gravity is through the Wald formula \cite{Wald}. 
Recently, it has been proposed  by A. Sen that the Wald formula for a specific class of extremal black holes in higher derivative 
gravity can be written  in terms of  the entropy function \cite{Sen}. 
The entropy function for the extremal black holes that 
their near horizon is $AdS_2\times S^{D-2}$ is defined by integrating the 
Lagrangian density over $S^{D-2}$ for a general $AdS_2\times S^{D-2}$ 
background characterized by the sizes of $AdS_2$ and $S^{D-2}$,  and taking 
the Legendre transform  of the resulting function with respect to the 
parameters labeling the electric fields. The result is  a function of   
moduli scalar fields as well as the sizes of $AdS_2$ and $S^{D-2}$. The  
values of the moduli fields and the sizes at near horizon are determined by extremizing 
the entropy function with respect to these  fields.  The entropy is then  given by 
the value of the entropy function at its extremum\footnote{In  above discussion, it has been  assumed 
that in the presence of higher derivative terms the near 
horizon geometry has the symmetry of $AdS_2\times S^{D-2}$. In the cases that the higher 
derivative corrections change this  symmetry, the near horizon solution can not be found by extremizing  the entropy function  and the Wald formula can not be written in terms of the entropy function. 
In those cases one should solve the differential equations of motion to find the near horizon solution and then  use the free energy method \cite{Hawking:1982dh} or the  Wald formula \cite{Wald} to calculate the 
entropy.}.   Using this method the near horizon solution and the entropy of some extremal black 
holes in the presence of higher derivative terms have been found in \cite{Sen},\cite{entfunc},\cite{Ghodsi:2006cd}.

The horizon in the extremal black hole that its near horizon solution has symmetry of $AdS_2\times S^{D-2}$ is an attractor, \ie the physical distance between an arbitrary point and the horizon is infinite \cite{Kallosh:2006bt}. In this case, the values of scalar fields at the near horizon are independent of the values of these fields at infinity. Hence, one expects the near horizon values of these fields to be given by some algebraic equations, \ie the equations that one finds by extremizing the entropy function \cite{Sen}. We will show in this paper that in some non-extremal (near extremal)  cases, even though the throat is not infinite, the near horizon values of the scalar fields are independent of the values of these fields at infinity, \ie they are given by some algebraic equations.    

It has been shown in \cite{mgag} 
that the entropy function has a saddle point at the near horizon of extremal 
black holes. This may indicates that  the entropy function formalism should 
not be specific to the extremal black holes.  In fact, it has been 
shown in \cite{mgag},\cite{Garousi:2007nn},\cite{Cai:2007ik} that the entropy function formalism works for 
 non-extremal black hole/branes at the supergravity level. However, the higher derivative terms in many cases change the symmetry of the tree level solutions, so one can not find the near horizon solution in these cases using the entropy function formalism, as in the extremal cases.  
 
 In this paper we would like to show that the higher derivative correction to the near horizon solution of the non-extremal (near extremal) $D1D5P$  can be calculated using the entropy function formalism and the Wald formula can be written in terms of the entropy function. We do this by explicitly solving the differential equations of motion and comparing the result with the near horizon solution that one finds using the entropy function formalism. Moreover, we will show that the entropy that one finds  from the free energy method is the same as the entropy that one finds by equating the Wald formula with the entropy function, as in the  extremal cases.

An outline of the paper is as follows. In section 2, we review the construction of near horizon solution of  non-extremal 
$D1D5P$. 
In sections 3, we add the higher derivative $R^4$ terms to the supergravity. We find the higher derivative correction to the near horizon solution by explicitly solving the differential equations of motion in section 3.1. In section 3.2, we find the same near horizon solution using the entropy function formalism. In section 4, we study  the entropy of $D1D5P$ system. In section 4.1, we calculate the entropy using the free energy method, and in section 4.2, we calculate the entropy by equating the Wald formula with the entropy function at its extremum. The results in both cases are the same.

\section{Review of  non-extremal $D1D5P$ solution}
In this section we review the non-extremal $D1D5P$  solution
of the effective action of type II string theory. The two-derivatives effective 
action in the string frame is given by
\beqa 
S=\frac{1}{16\pi G_{10}}\int d^{10}x\,\sqrt{-g}\bigg\{
e^{-2\phi}\left(R+4(\prt\phi)^2-\frac{1}{12}H_{(3)}^2\right)
-\frac{1}{2}\sum\frac{1}{n!}F_{(n)}^2+\cdots\bigg\}\,,\labell{tree}
\eeqa 
where $\phi$ is the dilaton, $H_{(3)}$ is NS-NS 3-form field
strength, and $F_{(n)}$ is the electric R-R n-form field strength where 
$n=1,3,5$ for IIB and $n=2,4$ for type IIA theory.  In above equation, dots 
represent Fermionic terms in which we are not interested. The effective action 
includes a Chern-Simons term which is zero for the $D1D5P$  solution. Moreover, for this solution $F_{(n)}=dC_{(n-1)}$. 
The 5-form field
strength tensor is self-dual, hence, it is not described by the
above simple action. It is sufficient to adopt the
above action for deriving the equations of motion, and impose the
self-duality by hand.

The non-extremal $D1D5P$ solution of the IIB effective action when 
$D1$-branes are along the compact $(z)$ direction $(S^1)$, $D5$-branes are along the compact $(z,x_1,x_2,x_3,x_4)$ directions $(S^1\times T^4)$, the KK momentum  $P=N/R$ is along the $(z)$ direction, and the non-compact directions are $(r,\theta,\phi,\psi)$,  is given 
by the following, (see e.g. \cite{ksks}):
\beqa
ds^2_{10}&=&(f_1f_5)^{-\frac{1}{2}}\bigg(-dt^2+dz^2+K(\cosh\alpha_m dt-\sinh\alpha_m dz)^2\bigg)\nonumber\\
&+&f_1^{\frac{1}{2}} f_5^{-\frac{1}{2}} \sum_{i=1}^4dx_i^2\, 
+(f_1 f_5)^{\frac{1}{2}}\bigg({\frac{dr^2}{1-K}}+ r^2 d\Omega_3^2\bigg)\,,\quad e^{-2\phi}=f_1^{-1}f_5\,,\cr &&\cr
 C_{tz}&=&\coth\alpha_1\left(\frac{1}{f_1}-1\right)+\tanh\alpha_1\,\,,\quad
C_{tzx_1\cdots x_4}=\coth\alpha_5\left(\frac{1}{f_5}-1\right)+\tanh\alpha_5\,,\nonumber           
\eeqa
where we have set the string coupling at infinity to be $g_s=1$. In above,
\beqa
K(r)=\frac{r_H^2}{r^2}\,,\quad  f_1(r)=1+\frac{r_H^2\sinh^2\alpha_1}{r^2}\,,\quad f_5(r)=1+\frac{r_H^2\sinh^2\alpha_5}{r^2}\,.
\eeqa
The three conserved charges are
\beq
Q_1=\frac{Vr_H^2\sinh(2\alpha_1)}{2}\,, \quad Q_5=\frac{r_H^2\sinh(2\alpha_5)}{2}\,,\quad N=\frac{R^2Vr_H^2\sinh(2\alpha_m)}{2}\,.
\eeq
The near horizon solution can be found by taking the following limit:
\beqa
r^2\ll  r_H^2 \sinh^2\alpha_1,_5\,,
\eeqa
 In this limit $\alpha_1$ and $\alpha_5$ are very large so  $\sinh\alpha_1,_5\approx \cosh\alpha_1,_5$. In terms of the new coordinate
\beq
\rho^2\equiv r^2+r_H^2 \sinh^2\alpha_m\,,
\eeq
the near horizon solution is 
\beqa
ds^2&=&-\frac{(\rho^2-\rho_+^2)(\rho^2-\rho_-^2)}{\lambda^2 \rho^2} dt^2+ \frac{\lambda^2 \rho^2}{(\rho^2-\rho_+^2)(\rho^2-\rho_-^2)}d\rho^2\labell{15p0}\\
&&+\frac{\rho^2}{\lambda^2}(dz-\frac{\rho_+\rho_-}{ \rho^2}dt)^2+
\lambda^2d\Omega_3^2+\frac{r_1}{r_5}\sum_{i=1}^4dx_i^2\,,\nonumber\\
e^{-2\phi}&=&\left(\frac{r_5}{r_1}\right)^2\,,\quad F_{t\rho z}=\frac{2\rho}{r^2_1}\,,\quad F_{t\rho zx_1\cdots x_4}=\frac{2\rho}{r^2_5}\,.
\eeqa
where
\beqa
r^2_{1,5}\equiv r_H^2 \sinh^2\alpha_1,_5\,,\quad\rho_+ \equiv r_H \cosh\alpha_m\,,& \rho_-\equiv r_H \sinh\alpha_m\,,& \lambda^2\equiv r_1r_5\,.
\eeqa
The above metric is a direct product of $ S^3\times T^4$ and the BTZ black hole \cite{mb} upon rescaling the coordinates.  The horizon $\rho=\rho_+$ in above solution is not attractor. The physical distance between an arbitrary point and horizon is 
\beqa
\int_{\rho_+}^{\rho}\frac{\lambda\rho d\rho}{\sqrt{(\rho^2-\rho_+^2)(\rho^2-\rho_-^2)}}&=&\frac{\lambda}{2}\ln\left(\frac{\rho^2-\frac{1}{2}(\rho_+^2+\rho_-^2)+\sqrt{(\rho^2-\rho_+^2)(\rho^2-\rho_-^2)}}{\frac{1}{2}(\rho_+^2-\rho_-^2)}\right)\,,\nonumber
\eeqa
which is finite. For the extremal case which  corresponds to $\rho_+=\rho_-$, the distance is infinite. For the near extremal case in which we are interested, however, the distance can be made as large as we want by sending $\rho_+\rightarrow \rho_-$. So one expects the asymptotic region to be decoupled from the near horizon region.

The higher derivative corrections to the supergravity action \reef{tree} may modifies the  near horizon solution \reef{15p0}. In general, they  have field redefinition freedom \cite{AAT,AAT1}, so one may choose different scheme  for the higher derivative terms. It has been argued in \cite{Gubser:1998nz} that the scheme in which the  corrections are written in terms of the 6-dimensional Weyl tensor, the near horizon solution \reef{15p0} is not modified so it may be the reason behind the equality of  the supergravity entropy  and  the entropy  from counting the degrees of freedom  for the non-extremal case \cite{Callan:1996dv}.  In the scheme that $R^4$ corrections are written in terms of 10-dimensional Weyl tensor, however, the solution \reef{15p0} is  modified which may indicate that the corrections associated  with the Ramond-Ramond  field have nontrivial contribution  to this solution in 10-dimensions. We will find the $R^4$  correction in the next section.  

\section{$R^4$ correction}
In this section we are going to  consider  the  string correction $\alpha ^{\prime 3}R^4$  to the supergravity action. The  correction in  the scheme that gravity is written in terms of the 10-dimensional Weyl tensors is \cite{higher}
\beqa
S=\frac{1}{16\pi G_{10}}\int d^{10}x\,\sqrt{-g}\bigg\{ {\cal L}^{tree}+
e^{-2\phi}\left(\gamma W\right)\bigg\}\,,\labell{action} 
\eeqa
where ${\cal L}^{tree}$ is given in \reef{tree}, $\gamma=\frac18\zeta(3)(\al')^3$ and $W$  in terms 
of the Weyl tensor is
\beqa
W=C^{hmnk}C_{pmnq}{C_h}^{rsp}{C^q}_{rsk}+\frac12 C^{hkmn}C_{pqmn}{C_h}^{rsp}{C^q}_{rsk}\labell{W}\,.
\eeqa
Using the above correction to the supergravity, one can find its effect on the non-extremal solution \reef{15p0}. This can be done by solving the differential equations of motion that we are going to do in the next section or by using the entropy function formalism that we will do in section 3.2. 

\subsection{Correction via solving differential E.O.M.}

We are going to work in Euclidean space in this section. In order to find a solution in the presence of higher derivative terms, one should make an ansatz for the solution and then find the unknown functions in the ansatz by solving the differential equations of motion. We consider the following ansatz for the solution:
\beqa
{ds^2}&=&a(\rho)\bigg(\frac{(\rho^2-\rho_+^2)(\rho^2-\rho_-^2)}{\lambda^2 \rho^2} d\tau^2+ \frac{\lambda^2 \rho^2}{(\rho^2-\rho_+^2)(\rho^2-\rho_-^2)}d\rho^2+\nonumber\\
&+&\frac{\rho^2}{\lambda^2}(dz-i\frac{\rho_+\rho_-}{ \rho^2}d\tau)^2\bigg)+b(\rho)\left({\lambda^2}d\Omega_3^2+(\frac{r_1}{r_5})\sum_{i=1}^4dx_i^2\right)\,, \labell{15p}\\
e^{-2\phi}&=&u(\rho)\,,\quad F_{\tau\rho z}\,=\,\frac{2i\rho}{r_1^2}\frac{a^{3/2}(\rho)}{b^{7/2}(\rho)}\,,\quad F_{\tau\rho z x_1\cdots x_4}\,=\,\frac{2i\rho}{r_5^2}a^{3/2}(\rho)b^{1/2}(\rho)\,,\nonumber
\eeqa
where $a(\rho),\, b(\rho)$ and $u(\rho)$  are the scalar fields. We have assumed the RR charges are not modified by the higher derivative correction. With the above ansatz, the Euclidean action becomes
\beq
{I}=-\frac{1}{16 \pi G_{10}}{\int d^9x} {\int d\rho }\left[ {\ell}{(a,a^{'},a^{''},...,u,u^{'},u^{''})+\gamma { \omega}{(a,a^{'},a^{''},...,u,u^{'},u^{''})}}\right]\,, 
\eeq
where $\ell$ and $\omega$ are
\beq
{\ell}=\sqrt{g} \left[ u(\rho) R -\frac 12 \frac{F_{(3)}^2}{3!}-\frac 12 \frac{F_{(7)}^2}{7!}\right]\,,\qquad \omega=\sqrt{g} u(\rho){W}\,,
\eeq
and 
\beqa
\sqrt{g}&=&\frac{\rho a^{3/2}(\rho)b^{7/2}(\rho) r_1^3}{r_5}\,,\quad \frac{F^2_{(3)}}{3!}\,=\,\frac{4r_5}{r_1^3b^7(\rho)}\,,\quad \frac{F^2_{(7)}}{7!}\,=\,\frac{4r_5}{r_1^3b^7(\rho)}\,.
\eeqa
The Euler-Lagrange equation for the scalar field $a(\rho)$ which follows from the above action is given by
\beqa
\frac {\partial \ell}{\partial a}-\frac{d}{d\rho}\frac {\partial \ell}{\partial a^{'}}+\frac{d^2}{d\rho^2}\frac{\partial ^2 \ell}{\partial a^{''}}=-\gamma \left(\frac {\partial \omega}{\partial a}-\frac{d}{d\rho}\frac {\partial \omega}{\partial a^{'}}+\frac{d^2}{d\rho^2}\frac{\partial ^2 \omega}{\partial a^{''}}\right)\,,\labell{diff}
\eeqa 
and similarly  for  $b(\rho)$ and $u(\rho)$. These differential equations are valid only to first order of $\gamma$, so one has to solve them perturbatively. At the zeroth order of $\gamma$, the solution is \reef{15p}, \ie $a=b=1,\, u=(r_5/r_1)^2$. At the first order of $\gamma$, the solution must be in the following form:
\beq
a(\rho)=1+\gamma a_p(\rho)\,,\qquad b(\rho)=1+\gamma b_p(\rho)\,,\qquad u(\rho)=\left(\frac{r_5}{r_1}\right)^2(1+\gamma u_p(\rho))\,.
\eeq
Inserting them in the differential equations of motion, one  finds the following equations for the scalars $a_p$, $b_p$, $u_p$, respectively: 
\beqa
\bigg(\rho{(\rho^2-\rho_+^2)(\rho^2-\rho_-^2)}\bigg)\bigg(\frac17{a_p}''+{b_p}''+\frac27{u_p}''\bigg)+\bigg({3\rho^4-\rho^2\rho_+^2-\rho^2\rho_-^2-\rho_+^2\rho_-^2} \bigg)\nonumber\\\bigg({\frac17{a_p}'+{b_p}'+\frac27{u_p}'}\bigg)-\bigg(\frac37\rho^3(a_p+7b_p+2u_p)\bigg)=-\frac{9}{28}\rho^3Q_1^{-3/2}Q_5^{-3/2}\,,\nonumber
\eeqa
\beqa
\bigg(\rho{(\rho^2-\rho_+^2)(\rho^2-\rho_-^2)}\bigg)\bigg({a_p}''+3{b_p}''+{u_p}''\bigg)+\bigg({3\rho^4-\rho^2\rho_+^2-\rho^2\rho_-^2-\rho_+^2\rho_-^2} \bigg)\nonumber\\\bigg({{a_p}'+3{b_p}'+{u_p}'}\bigg)-\bigg(3\rho^3(a_p-\frac{61}{21}b_p-\frac27u_p)\bigg)=\frac{27}{28}\rho^3Q_1^{-3/2}Q_5^{-3/2}\,,\nonumber
\eeqa
\beqa
\bigg(\rho{(\rho^2-\rho_+^2)(\rho^2-\rho_-^2)}\bigg)\bigg(\frac27{a_p}''+{b_p}''\bigg)+\bigg({3\rho^4-\rho^2\rho_+^2-\rho^2\rho_-^2-\rho_+^2\rho_-^2} \bigg)\nonumber\\\bigg({\frac27{a_p}'+{b_p}'}\bigg)+\bigg(\frac67\rho^3(-a_p+b_p)\bigg)=\frac{9}{14}\rho^3Q_1^{-3/2}Q_5^{-3/2}\,.
\eeqa
The above differential equations should give correction to  the near horizon geometry \reef{15p}. Similar equations have been found in \cite{Gubser:1998nz} for the correction to the non-extremal D3-branes and in \cite{Garousi:2008ik} for the non-extremal M2-branes. In those cases, using the boundary condition at the horizon, one finds that  the solution is a power law solution. However,  the above equations have only constant solutions  with the following values:   
\beqa
a(\rho)&=&1-\gamma\frac{51}{32 r_1^{3}r_5^{3}}\,,\notag\\b(\rho)&=&1-\gamma\frac{27}{32 r_1^{3}r_5^{3}}\,,\notag\\u(\rho)&=&\left(\frac{r_5}{r_1}\right)^2(1+\gamma\frac{33}{8 r_1^{3}r_5^{3}})\,.\labell{pert}
\eeqa
This indicates that the near horizon geometry in the presence of the higher derivative terms is a direct product of $ S^3\times T^4$ and the BTZ black hole, as in the tree level. In \cite{Cai:2007ik},  it has been shown that  the tree level solution \reef{15p} is consistent with  the entropy function formalism. The above result indicates that this consistency should be valid even in the presence of   the higher derivative terms. Hence, the  correction to the non-extremal solution \reef{15p}  should  be also found by using the entropy function formalism that we are going to do in the next section. We note however that there are many cases that the entropy function formalism works only at the tree level, \eg non-extremal D3, M2, M5 solutions \cite{mgag}.

\subsection{Correction via entropy function formalism}

We are going to work in Minkowski  space in this section. In order to find solution in the entropy function formalism, one should consider a general background with the same symmetry as the symmetry of the tree level solution, \ie 
\beqa
{ds^2}&=&v_1\bigg(-\frac{(\rho^2-\rho_+^2)(\rho^2-\rho_-^2)}{\lambda^2 \rho^2} dt^2+ \frac{\lambda^2 \rho^2}{(\rho^2-\rho_+^2)(\rho^2-\rho_-^2)}d\rho^2\labell{15pp}\\
&+&\frac{\rho^2}{\lambda^2}(dz-\frac{\rho_+\rho_-}{ \rho^2}dt)^2\bigg)+v_2\left({\lambda^2}d\Omega_3^2+(\frac{r_1}{r_5})\sum_{i=1}^4dx_i^2\right)\,,\nonumber\\
e^{-2\phi}&=&u_s\,,\quad F_{t\rho z}\,=\,\frac{2\rho}{r_1^2}\frac{v_1^{3/2}}{v_2^{7/2}}=e_1\,,\quad F_{t\rho z x_1\cdots x_4}\,=\,\frac{2\rho}{r_5^2}v_1^{3/2}v_2^{1/2}=e_2\,,\nonumber
\eeqa
where the scalars $v_1,\,v_2,\, u_s$ are constant. The algebraic equations from which  these constant parameters can be  found are given by extremizing   the entropy function. The entropy function, on the other hand, is defined by taking  the Legendre transform of function $f(v_1,v_2,u_s,e_1,e_2,\rho)=\int_Hdx_H\sqrt{-g}L$ with respect to the electric fields $e_1,\, e_2$, and dividing the result  by $\rho$. That is
\beqa
F(v_1,v_2,u_s)&=&\frac{1}{\rho}\left(e_1\frac{\prt f}{\prt e_1}+e_2\frac{\prt f}{\prt e_2}-f\right)\,.
\eeqa
Using the Lagrangian in \reef{action}, one finds the following entropy function:
\beqa
 F(v_1,v_2,u_s)&=&\frac{V_1V_3V_4}{16\pi G_{10}}\bigg(6u_s \frac{r^2_1}{r^2_5} v_1^{1/2}v_2^{5/2}(v_2-v_1)+\frac{2v_1^{3/2}}{v_2^{7/2}}+2v_1^{3/2} v_2^{1/2}\nonumber\\
&-&\gamma u_s \frac{r_1^{3}}{r_5^{}}v_1^{3/2}v_2^{7/2}
\frac{105 v_2^4-60 v_1^3 v_2+ 54 v_1^2 v_2^2-60 v_1 v_2^{3}+ 105 v_1^{4}}{32r_1^4 r_5^4 v_1^4 v_2^4}\bigg)\,.\labell{Free}
\eeqa
Considering the following perturbative solution: 
\beq
v_1=1+\gamma x \,,\qquad v_2=1+\gamma y \,,\qquad u_s=\left(\frac{r_5}{r_1}\right)^2(1+\gamma z)\,,
\eeq
one finds the following equations 
\beqa
\frac{\prt F}{\prt u_s}&=&0  \,\qquad{\rightarrow} \,\qquad 6(x-y)=\frac{9}{2(r_1r_5)^{3}}\,,\nonumber\\
\,\qquad \frac{\prt F}{\prt v_1}&=&0  \,\qquad{\rightarrow} \,\qquad 28y+4x+8z=\frac{3}{(r_1r_5)^{3}}\,,\nonumber\\
\frac{\prt F}{\prt v_2}&=&0  \,\qquad{\rightarrow} \,\qquad -244y+84x-24z=-\frac{27}{(r_1r_5)^{3}}\,.
\eeqa
The solution to these consistent equations is
\beq
v_1=1-\gamma \frac{51}{32(r_1r_5)^{3}}\,,\quad v_2=1-\gamma \frac{27}{32(r_1r_5)^{3}}\,,\quad u_s=\left(\frac{r_5}{r_1}\right)^2(1+\gamma \frac{33}{8(r_1r_5)^{3}})\,,\labell{solution}
\eeq
which is exactly the same as the solution in \reef{pert}. Therefore, even though this system is non-extremal, the entropy function formalism and the differential equations of motion yield the same result for the near horizon background in the presence of the higher derivative terms. We now turn to the calculation of  the entropy of this system.

\section{Entropy of nonextremal $D1D5P$ system}

 In the presence of higher derivative terms, the entropy can be calculated  either from the free energy or from the Wald formula. For the systems that the entropy function formalism can be used to find the near horizon solution, \eg the non-extremal $D1D5P$ solution, the Wald formula can be written in terms of the entropy function which is an efficient way to calculate the entropy in the presence of higher derivative terms. In the next section we calculate the entropy using the free energy method \cite{Hawking:1982dh}, and in the section 4.2 we calculate the entropy from the Wald formula.
 
\subsection{Entropy from free energy}
 Following \cite{Hawking:1982dh}, one can identify the free energy of the theory with the Euclidean gravitational action, $I$, times the temperature, $T$, i.e.
\beq
I=\beta \cF\,,
\eeq  
where $\beta=\frac{1}{T}$. 
The calculation of the Euclidean action is divergent at large distances, $\rho_{max}$, and requires a subtraction. The integral must be regulated by subtracting off its zero entropy limit, i.e.
\beq
\cF=\lim_{\rho_{max}\rightarrow\infty}\frac{I-I_{0}}{\beta}\,,
\eeq
where $I_0$ is the zero entropy limit in which the periodicity of the Euclidean time is defined by $\beta_0$. One must adjust $\beta_0$ so that
the geometry at $\rho=\rho_{max}$ is the same in the two cases, i.e. the black hole and its zero entropy limit. This can be done by equating the circumference of the Euclidean time in two cases.
%\beq
%\int_0^{\beta_0}\sqrt{g^{{\rm (zero\, temperature)}}_{\tau\tau}}|_{\rho_{max}}d\tau=\int_0^\beta\sqrt{g^{{\rm (black\, %hole)}}_{\tau\tau}}|_{\rho_{max}}d\tau.
%\eeq
Having  $\cF$, the entropy in terms of the free energy is then given by $S=-\frac{\partial \cF}{\partial T}$.

Let us start from the black hole action which will be noted by $I_{BH}$
\beq
I_{BH}=-\frac{1}{16\pi G_{10}}\int d^{10}x\,\sqrt{g}\ {\cal L}\,,
\eeq
where ${\cal L}$ is given in \reef{action}. Inserting the solution \reef{15p} in which $a(\rho),\,b(\rho), \, u(\rho)$ are given in \reef{pert}, one finds  
\beqa
I_{BH}&=&\frac{1}{4\pi G_{10}}\int^{\beta}_{0}d\tau\,\underbrace{\int dz}_{V_1}\,\underbrace{\int d{\Omega_3}}_{V_3}\,\underbrace{\int dx_1...dx_4}_{V_4}\int^{\rho_{max}}_{\rho_+}\left( 1-\gamma\frac{9 }{8r_1^{3}r_5^{3}}+O(\gamma^2)\right)\rho d\rho\nonumber\\
&=&\frac{1}{8\pi G_{10}}V_1V_3V_4\beta\left( 1-\gamma\frac{9 }{8r_1^{3}r_5^{3}}+O(\gamma^2)\right)(\rho^2_{max}-\rho^2_+)\,,
\eeqa
where $\rho_{max}$ is a cutoff at large distances.
The above expression  is divergent at large distances and must be regulated by subtracting off its zero entropy limit, $I_{AdS}$. It is given by 
\beqa
I_{AdS}=&=&\frac{1}{4\pi G_{10}}\int^{\beta_0}_{0}d\tau\,{\int dz}\,{\int d{\Omega_3}}\,{\int dx_1...dx_4}\int^{\rho_{max}}_{0}\left( 1-\gamma\frac{9 }{8r_1^{3}r_5^{3}}+O(\gamma^2)\right)\rho d\rho\,,\nonumber\\
&=&\frac{1}{8\pi G_{10}}V_1V_3V_4\beta_0\left( 1-\gamma\frac{9 }{8r_1^{3}r_5^{3}}+O(\gamma^2)\right)(\rho^2_{max})\,.
\eeqa
The relation between $\beta_0$ and $\beta$ is
\beq
\beta_0=\beta \sqrt{\frac{(\rho^2-\rho_{+}^2)(\rho^2-\rho_{-}^2)}{\rho^4}}\bigg |_{\rho=\rho_{max}}\cong \beta\bigg ({1-\frac{(\rho_{+}^2+\rho_{-}^2)}{2\rho_{max}^2}}\bigg)\,,
\eeq
which comes from the fact that the geometry of the hypersurface $\rho=\rho_{max}$ must be the same for both cases \cite{Hawking:1982dh}. 

Taking the limit $\rho_{max} \rightarrow \infty$  of the subtraction of  $I_{BH}$ and $I_{AdS}$, one finds the free energy in terms of $r_H$ to be 
\beq
\cF= \lim_{\rho_{max} \rightarrow \infty} \bigg(\frac{I_{BH}-I_{AdS}}{\beta}\bigg)=-\frac{1}{16\pi G_{10}}{V_1V_3 V_4}\left( 1-\gamma\frac{9 }{8r_1^{3}r_5^{3}}+O(\gamma^2)\right)r_H^2\,.
\eeq 
To write it in terms of temperature, one calculates the  surface gravity by KK reduction to 9-dimension (see \eg \cite{Peet:2000hn}) which has diagonal metric, \ie
\beq
\kappa = 2\pi T=\sqrt{G^{\rho\rho}} {\frac {d}{d\rho}}\sqrt{G_{\tau \tau}}\ \bigg|_{Horizon}\,,
\eeq
where
\beqa
G_{\tau \tau}&=&g_{\tau \tau}-\frac{g_{\tau z}^2}{g_{z z}},\quad G^{\rho\rho}\,=\,g^{\rho\rho}\,.
\eeqa
One finds the temperature to be
\beqa
T&=&\frac{(\rho_{+}^2-\rho_{-}^2)}{2\pi\lambda^2\rho_{+}}=\frac {1}{2\pi r_H\cosh(\alpha_1)\cosh(\alpha_5)\cosh(\alpha_m)}\,,\nonumber
\eeqa
Using the fact that the number of D1 and D5 branes, $N_1,\, N_5$, and the boost parameter $\alpha_m$ are independent of temperature, one finds the following linear relation between temperature and $r_H$: 
\beqa
T&=&\frac{\sqrt{V}}{2\pi \ell_s^4\sqrt{N_1N_5}\cosh\alpha_m}r_H\,,
\eeqa
where  we have used
\beq
N_1=\frac{V r_H^2}{2 l_s^6}\sinh(2\alpha_1)\,,\quad N_5=\frac{r_H^2}{2 l_s^2}\sinh(2\alpha_5)\,, 
\eeq
and the fact that in the near horizon region  $\sinh(\alpha_{1,5})\approx\cosh(\alpha_{1,5})$. The entropy $S=-\frac{\prt \cF}{\prt T}$ becomes
\beqa
S&=&\frac{1}{8\pi G_{10}}{V_1V_3 V_4}\left( 1-\gamma\frac{9 }{8r_1^{3}r_5^{3}}+O(\gamma^2)\right)\frac{2\pi\ell_s^4\sqrt{N_1N_5}r_H\cosh\alpha_m}{\sqrt{V}}\,.
\eeqa
It is convenient to write the entropy in terms of the left and right KK momenta,  $N_L$ and $N_R$ which are defined as 
\beqa
N_{L,R}=\frac{VR_z^2r_H^2}{4l_s^8}\exp({\pm2\alpha_m})\,,\labell{nlnr}
\eeqa
Using above relations, one finds the entropy to be
\beq
S=2\pi \sqrt{N_1N_5}(\sqrt{N_L}+\sqrt{N_R})\left( 1-\gamma\frac{9 }{8r_1^{3}r_5^{3}}+O(\gamma^2)\right)\,.\labell{entropy1}
\eeq
where we have also used the relevant formulas for the volume of the circle, 3-sphere and the volume of 4-torus as well as the the 10-dimensional Newton constant, \ie   
\beqa
V_1=2\pi R_z\,,\qquad V_3=2\pi^2\,,\qquad V_4=(2\pi)^4V\,,\qquad{G_{10}} ={8 {\pi ^6}  {\ell_s}^8}\,,\labell{vs}
\eeqa
The first term in \reef{entropy1} is the supergravity result (see \eg \cite{Peet:2000hn}) and the second term is the higher derivative correction. In the next section we calculate the entropy using the entropy function formalism. 
 
\subsection{Entropy from entropy function}
 
Entropy in a higher derivative theory can also be calculated from the  Wald  formula \cite{Wald}
\beqa
S_{BH}=4\pi\int_H dx_H{\sqrt{-g_H}} {\frac{\prt L}{\prt R_{\mu\nu\lambda\rho}}g_{\mu\nu}^\bot g_{\nu\rho}^\bot}\,,\labell{wald}
\eeqa
where $L$ is the Lagrangian density and $g_{\mu\nu}^\bot$ denotes the metric projection onto subspace orthogonal to the horizon.
It has been shown in \cite{Sen} that for extremal black holes that  the near horizon geometry can be calculated using  the entropy function formalism,   the Wald formula  is proportional to  the entropy function. We have seen that the correction to the non-extremal solution \reef{15p} can also be calculated using the entropy function formalism. Hence, one expects  that the Wald formula in this case also is  proportional to the entropy function. The constant of the proportionality can be fixed by comparing it  with the entropy at the supergravity level. That is  
\beqa
S_{BH}&=& \frac{\pi\ell_s^4\sqrt{N_1N_5}\rho_+}{\sqrt{V}}F\,,\labell{SBH1}
\eeqa
where $F$ is the entropy function. One can easily check that the above entropy is the same as the tree level entropy \reef{entropy1} after  inserting the tree level entropy function  \reef{Free} into it. The above formula can also be found  directly from the Wald formula \cite{Ghodsi:2006cd,Garousi:2007nn,Cai:2007ik}. 

Now, inserting the solution \reef{solution} into the entropy function \reef{Free}, one finds the entropy function at its extremum  to be
\beqa
F&=&\frac{1}{8\pi G_{10}}{V_1V_3 V_4}\left( 1-\gamma\frac{9 }{8r_1^{3}r_5^{3}}+O(\gamma^2)\right)
\eeqa
After inserting this into \reef{SBH1} and using \reef{nlnr} and \reef{vs}, one finds exactly the entropy in \reef{entropy1}. This confirms that the Wald formula \reef{wald} for the non-extremal $D1D5P$ solution  in the presence of higher derivative terms is proportional to the entropy function. For extremal case, \ie $N_R=0$, the entropy has been found in \cite{Ghodsi:2006cd}, however, the correction is different from the one in \reef{entropy1}. This is related to the fact that the scheme for the higher derivative terms in \cite{Ghodsi:2006cd} is different from the scheme that we have chosen in \reef{W}. When there is no KK momentum, \ie $N_R=N_L$, the entropy \reef{entropy1} is the same as  the entropy that has been found in \cite{Garousi:2007nn}.

We have done the same calculation for the non-extremal $D2D6NS5P$ solution and found that  the correction to the tree level solution can be calculated either by solving the differential equations of motion or by using the entropy function formalism. In this case also the Wald formula is proportional to the entropy function and it is equal to the entropy that one finds using the free energy method.

The reason that the entropy function formalism works for the non-extremal (near extremal) $D1D5P$ and $D2D6NS5P$ cases may be related to the fact that  the near horizon geometry of  these solutions have a throat. Even though the physical length of the throat is finite for non-extremal cases, the throat can be made as long as we want for near extremal cases. Hence, the scalar fields at the horizon are independent of the values of these fields at infinity.  The tree level solutions  of non-extremal (near extremal) $D3,\, M_2$ and $ M5$ at the near horizon also have throat geometries, however, the modified solutions in the presence of the higher derivative correction have no longer the throat. That is why the entropy function formalism works for these systems only at tree level. For non-extremal $D1D5P$ and $D2D6NS5P$ cases, however, the higher derivative corrections keep the tree level  throat. Hence, the entropy function formalism works even in the presence of the higher derivative terms. 

In fact the entropy function formalism for extremal cases \cite{Sen} works in above sense. That is, the original Wald formula  for black hole entropy  holds for non-extremal black holes, and in applying this result to extremal case one must define the entropy of an extremal black hole to be  the limit of the entropy of the associated  non-extremal black hole in which the non-extremal parameter goes to zero. For example, the entropy of extremal $D1D5P$ is given by the entropy of non-extremal $D1D5P$ \reef{entropy1} in which $N_R\rightarrow 0$. Hence,  
 one may expect the entropy function formalism works for any near extremal solution which has throat at the near horizon region, and  entropy to leading order of the non-extremal parameter can be found using the entropy function formalism.

\end{document}